\documentclass[preprints,article,accept,moreauthors,10pt,a4paper]{mdpi} 

\firstpage{1} 
\pubvolume{xx}
\issuenum{1}
\articlenumber{5}
\pubyear{2019}
\copyrightyear{2019}
\history{Received: date; Accepted: date; Published: date}

\Title{Molecular Diversity and Network Complexity in Growing Protocells}

\Author{Atsushi Kamimura $^{1,*}$ and Kunihiko Kaneko $^{1,2,*}$}

\AuthorNames{Atsushi Kamimura and Kunihiko Kaneko}

\address{%
$^{1}$ \quad Research Center for Complex Systems Biology, Universal Biology Institute, The University of Tokyo, 3-8-1, Komaba, Meguro-ku, Tokyo 153-8902, Japan \\
$^{2}$ \quad Department of Basic Science, Graduate School of Arts and Sciences, The University of Tokyo, 3-8-1, Komaba, Meguro-ku, Tokyo 153-8902, Japan}

\corres{Correspondence: kamimura@complex.c.u-tokyo.ac.jp(AK); kaneko@complex.c.u-tokyo.ac.jp(KK)}

\abstract{A great variety of molecular components is encapsulated in cells. Each of these components is replicated for cell reproduction. To address an essential role of the huge diversity of cellular components, we study a model of protocells that convert resources into catalysts with the aid of a  catalytic reaction network. As the resources are limited, it is shown that diversity in intracellular components is increased to allow the use of diverse resources for cellular growth. Scaling relation is demonstrated between resource abundances and molecular diversity. We then study how the molecule species diversify and complex catalytic reaction networks develop through the evolutionary course. 
It is shown that molecule species first appear, at some generations, as parasitic ones that do not contribute to replication of other molecules.  
Later, the species turn to be host species that support the replication of other species. With this successive increase of host species, a complex joint network evolves. 
The present study sheds new light on the origin of molecular diversity and complex reaction network at the primitive stage of a cell.  }


\begin{document}


\section{Introduction}
\label{Introduction}

Diversity is one of the fascinating features of life. Diverse molecule species are encapsulated and coexist in cellular compartments. They are synthesized with the aid of catalysts for cell reproduction. Then, why do cells have so many components? 
The question arises because such a great diversity of molecule species is not a strict requirement for cell reproduction. 
Rather, it would not be fitted to realize a higher growth of a cell. 
In fact, a simple cell consisting of fewer components is generally expected to achieve a faster growth speed. This expectation is supported by several in vitro and in silico models. A replicator system with few components drives out a complex system with diverse components\cite{Mills217,Kacian3038,ichihashi2013darwinian,fontana1994arrival,ray:91}. 

For self-sustaining reproduction, a minimum level of diversity is required to form an autocatalytic set in collectively catalytic chemical reaction networks\cite{eigenhypercycle,dyson,kauffman1986autocatalytic,jain1998autocatalytic,GARD,furusawa2003zipf,kaneko2003recursiveness,KamimuraKaneko2010}. 
Still, diversification beyond the minimum requirement will decrease the fitness (growth rate). 
A minimum cell with essential components would achieve a higher reproduction rate than a complex cell with a huge diversity of molecules. Thus, diversity would be evolutionarily selected out.
The present cells, however, consist of a huge diversity of molecules. This issue on diversity along with cell reproduction, then has to be addressed generally also for protocells\cite{bedau2009protocells,szostak2001synthesizing,ganti2003principles,luisi2016emergence,ruiz2013prebiotic}. 

In considering that cells with fewer components have higher growth, one implicitly assumes that resources used for synthesizing each component are sufficiently supplied and are always abundant. By consuming the resources, cells with a minimum set of components increase their population. 
As the population increases, however, the resources actually get limited. 
When the resources are limited, cells with diverse components may have a potential to use different resources in the environment, which could help them keep the growth of cells. Then, the diversification of cellular components may be favorable. 
Still, if and how the diversification progresses remain elusive. 

Recently, we have considered a protocell model in which catalytic molecules are replicated from resources, catalyzed by each other \cite{KamimuraKaneko2015,KamimuraKaneko2016}. In this paper, we first review the main findings of the model. By taking the consumption of resources into account, protocells with diverse molecule species emerge as they can utilize a variety of resources for their own growth. Under a selection pressure for the cells to grow faster, diversification of the molecule species occurs when the resources are limited. We then elucidate how the number of molecule species increases with the decrease in the resource abundances. A scaling relation is derived between molecule diversity and resource uptake, as the optimum diversity to achieve the maximum growth speed. The growth speed is maximized by a trade-off between the utility of diverse resources and the concentration onto fewer species to increase the reaction rate.

In the model, the growth rate is maximized by optimizing the number of molecule species. 
How the cells diversify their molecule species, however, is not explored in the previous study. 
In the present paper, we investigate an evolutionary constraint for it. 
The question we address here is how a molecule emerged by mutation can be fixed and increases its population.

In a cell that grows and divides, the fixation of new molecule species is highly probable if replication of the species is catalyzed by the remaining species. Then, the diversification occurs by adding the species one by one to the existing catalytic network. As a result, the number of species increases in a cell, which constitutes a connected reaction network. 
In principle, such a complex network is not essential for high fitness (growth rate), but the evolutionary constraint selects such a connected network rather than disconnected ones. 

Through the evolutional pathway of diversification, we highlight the potential importance of parasitic molecule species. The diversity is increased by the appearance of parasitic molecule species first and then that parasitic to such parasitic molecule species. Later they turn to be host species with a further increase in the number of species. By this way, the ``core" reaction network is shifted from a simple to a complex one.

The paper is organized as follows. In section \ref{Model}, we introduce a simplified cell model consisting of a catalytic reaction network under multiple resources in the environment. First, we briefly provide a mathematical condition for the coexistence of multiple replicators under resource limitation in section \ref{Result1}. In section \ref{Result2}, we review the diversification of intracellular components by resource limitation and present a general scaling relation between molecular diversity and the uptake of resources. In section \ref{Result3}, we discuss an evolutionary constraint for cells to satisfy the growth and diversification of components simultaneously and point out the relevance of complex, connected catalytic reaction networks. In section \ref{discussion}, we summarize and discuss our results. Details of simulation methods are given in section \ref{methods}.
 
\section{Model}
\label{Model}



\begin{figure}
\centering
\includegraphics[width=\textwidth]{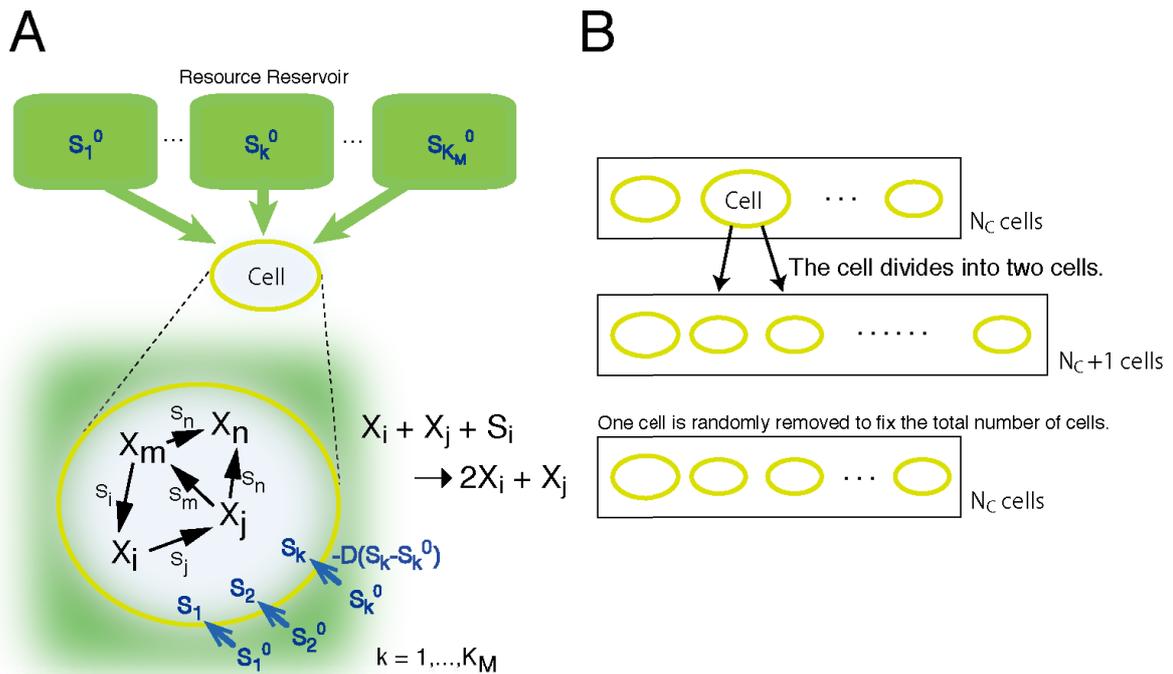}
\caption{Schematic representation of our model. Our model is composed of $N_C$ cells. (\textbf{A}) Each of the cells contains molecules $X_k$ and resources $S_k$ $(k = 1,...,K_M)$. The molecule species form a catalytic reaction network to replicate each of $X_k$. For example, with the aid of the catalytic molecule $X_j$, the molecule $X_i$ is replicated when at least one resource molecule $S_i$ is available in the cell. The resource $S_i$ is consumed to replicate $X_i$. Each cell takes up resources $S_k$ from the resource reservoir in the environment with the rate $D (S^0_k - S_k)$. The concentration of each $S_k^0$ in the environment is given by a random number $S^0_k \in [0,10]$ and is set fixed. The parameter $D$ controls the rate of the uptake. By replicating $X_k$ with the consumption of $S_k$, the number of molecules $X_k$ increases in each cell. 
(\textbf{B}) A cell divides when the total number of molecules exceeds a threshold of $N$. The content of the cell is randomly distributed into two daughter cells. At the same time, one cell is randomly removed from the system to fix the total number of cells at $N_C$. }
\label{fig1}
\end{figure}

We adopt a cell model which consists of $K_M$ species of molecules and resources [Figure \ref{fig1}A]. We denote each molecule and resource species by $X_i$ and $S_i$ $(i = 1,..,K_M)$, respectively. The molecules and resources are encapsulated in each of $N_C$ cells. Chemicals are well-mixed within a cell, so that sets of the amount of $X_i$ and $S_i$ for $i = 1,..., K_M$ determine the internal state of each cell. Some of the molecule species can have a null population. 
Inside each cell, the molecules $X_i$ replicate by consuming a corresponding resource $S_i$, with the aid of other molecules $X_j$ as  
\begin{equation}
X_i + X_j + S_i \rightarrow 2X_i + X_j.
\label{reaction}
\end{equation}
For the replication of $X_i$ by this reaction, one resource molecule of $S_i$ is needed, and the replication reaction does not occur if the number of $S_i$ is less than 1. The reaction coefficient is given by a catalytic activity $c_j$ of the molecule $X_j$. The activity $c_j$ is randomly determined and fixed as $c_j \in [0,1]$ for each $X_j$, throughout each set of simulations. Hence, a resource $S_i$ with highest $c_j$ is most efficient for the replication. 

At each replication, an error occurs with a probability $\mu$. This error corresponds to changes (replacement, insertion, or deletion) in the polymer sequence, which can alter the catalytic activity of the molecule. Here, for simplicity, for each replication of $X_i$, the molecule is replaced by a different molecule $X_k$ $(k = 1,..., K_M; k \neq i)$ with equal probability $\mu/(K_M-1)$.

Each cell takes up resources ($S_i$) from its respective resource reservoir. From the external reservoir, each resource $(S_i)$ is supplied into each cell with the rate $D(S^0_i - S_i)$. The coefficient $D$ controls the degree of the uptake rate because the resource supply is reduced by decreasing $D$. Here, the resources are supplied into each cell without competition among cells. We adopt this simplification since we focus on compositional diversification rather than cellular diversification. We carry out stochastic simulations of the model, as detailed in section \ref{methods}.

The catalytic relation between $X_i$ and $X_j$, i.e., the catalytic reaction network, is determined by a random assignment. For each pair of $X_i$ and $X_j$, the catalytic reaction path is assigned with probability $p$ (which was fixed at 0.1). Thus, each species has $p K_M$ reactions on average. The assignment excludes autocatalytic and direct mutually catalytic reactions. In other words, the species $X_i$ does not catalyze the replication of itself and that of molecules which directly catalyze the replication of $X_i$.

Once the catalytic reaction network is determined, it does not change throughout each simulation and is identical for all cells. Even if the underlying network is vast, each cell uses only a subset of the reaction pathways because both $X_i$ and $X_j$ must be present in the cell for the reaction (\ref{reaction}) to occur, whereas the number of molecules in a protocell is finite as will be given below. 

The cell divides into two when the total number of molecules in the cell exceeds a given threshold $N$ [Figure \ref{fig1}B]. The molecules and resources within the cell are randomly partitioned into two daughter cells. At the division event, one cell is randomly taken out from the system and removed, to fix the total number of cells at $N_C$. This leads to the selection of a protocell that can grow faster under a given resource condition.

\section{Diversification under resource limitation}
\label{Result}
\unskip
\subsection{Simple illustration of diversity transition}
\label{Result1}

Before investigating the above cell model with a mutually catalytic network, we briefly review a mathematical basis of diversification in an ensemble of simple replicators\cite{eigenhypercycle, szathmary1991simple, Kamimura_2018}.
Here, the mutually catalytic reaction in $\{ X_i \mid i = 1,..., K_M \}$ is not considered for mathematical simplicity. 
We consider only replication of a molecule $R_i$ by using itself as a template, by consuming a resource $S_i$. 
Here, the reaction is written as 
\begin{equation}
R_i + S_i \rightarrow 2R_i,
\end{equation}
where $R_i$ is a replicator $(i = 1,..., K_R)$ and $S_i$ is the corresponding resource needed for its replication.
By writing the concentration of $R_i$ as $\rho_i$$(i = 1,...,K_R)$, the dynamics is written as
\begin{align}
\frac{d\rho_i}{dt} &= a_i S_i \rho_i - \rho_i \phi, \label{r1} \\
\frac{dS_i}{dt} &= - a_i S_i \rho_i + D_r (S_i^0 - S_i),\label{r2}
\end{align}
where $a_i$ denotes the rate constant, and $\phi = \sum_j a_j S_j \rho_j$. The term with $\phi$ is introduced to fix the total population $\sum_j \rho_j = 1$. The parameter $D_r$ denotes the supply rate of the resource $S_i$ from the external environment which has a constant concentration $S^0_i$ for each resource $i$. 
We consider the case that replication rate of each molecule is not identical, i.e., $a_i S_i \neq a_j S_j$ for $i \neq j$.

From the steady state condition of Eq. (\ref{r2}), $dS_i/dt = 0$, one obtains
$\bar{S}_i = D_r S^0_i/(a_i \rho_i + D_r)$.
Thus, the value of $\bar{S}_i$ is written as 
\[ \bar{S}_i =\begin{cases}
S^0_i & (D_r \gg a_i \rho_i)\\
D_rS^0_i/a_i \rho_i & (D_r \ll a_i \rho_i).
\end{cases}
\]
By substituting $\bar{S}_i$ into Eq. (\ref{r1}), one obtains
\begin{align} 
\frac{d\rho_i}{dt} =\begin{cases}
a_i S^0_i \rho_i - \rho_i \phi & (D_r \gg a_i \rho_i) \\
D_r S^0_i - \rho_i \phi & (D_r \ll a_i \rho_i).
\label{r2_2}
\end{cases}
\end{align}
When $D_r \gg a_i \rho_i$, the steady state condition of Eq. (\ref{r2_2}) gives $\rho_i = 0$ or $a_i S^0_i = \sum_j a_j S^0_j \rho_j$.
If $\rho_i = 0$ for all $i$,  all the replicators were absent, and the condition $\sum_j \rho_j = 1$ could not be satisfied. At least for some replicators, the condition $a_i S^0_i = \sum_j a_j S^0_j \rho_j$ has to be satisfied.  
For the condition, $a_i S^0_i = \sum_j a_j S^0_j \rho_j$, the right-hand side is independent of the species $i$, whereas the left-hand side depends on the species $i$.
Therefore, the condition is satisfied only for a single species $i'$, because $a_i S_i \neq a_j S_j$ for $i \neq j$. All the molecule species except $i'$  do not exist, i.e., 
$\rho_{i'} = 1$ for a specific $i'$ and $\rho_j = 0$ for $j \neq i'$. 
Among these $K_R$ solutions, only such a case $i' = m$ such that $a_{m} S_{m}^0$ is the largest is stable. In other words, the species with the highest $a_m S_m^0$ outcompetes the others. Thus, the Darwinian selection occurs and the fastest replicator wins. 

On the other hand, when $D_r \ll a_i \rho_i$ in Eq. (\ref{r2_2}), the steady state condition gives $D_r S_i^0 = \rho_i \sum_j D_r S^0_j$.  
Here, coexistence is possible ($\rho_i = S^0_i/\sum_j S^0_j$). 
Because the resource supply is limited, any replicator cannot increase its population to outcompete the others. Thus, multiple replicators can coexist.

When the replicating molecules are encapsulated in growing cells, there are two distinct levels of replicating entities. 
At the molecule level, $R_i$ is a replicating molecule. At the cell population level, each cell is a replicator. 
In the cell model of section \ref{Model}, the replicating molecules are a set of molecules $\{ X_i \}$. 
In such a multi-level system, selections at both molecular and cellular levels have to be consistently satisfied. 
A fast-replicating entity wins when resources are abundant. 
Only when resources are limited, coexistence of multiple replicators is possible. As a result of the consistency of molecule replication and cell reproduction, the diversification of the cellular components occurs when the resources are limited.
In the next section, we will show that diversification also occurs for the mutually catalytic reaction in $\{ X_i \}$, besides the diversification for $R_i$.

\subsection{Negative scaling relation}
\label{Result2}

In our previous publication\cite{KamimuraKaneko2016}, we investigate how diversity in cellular composition changes with the uptake rate of the resources $D$ by numerical simulations of the model in section \ref{Model}. 

\begin{figure}
\centering
\includegraphics[width=14cm]{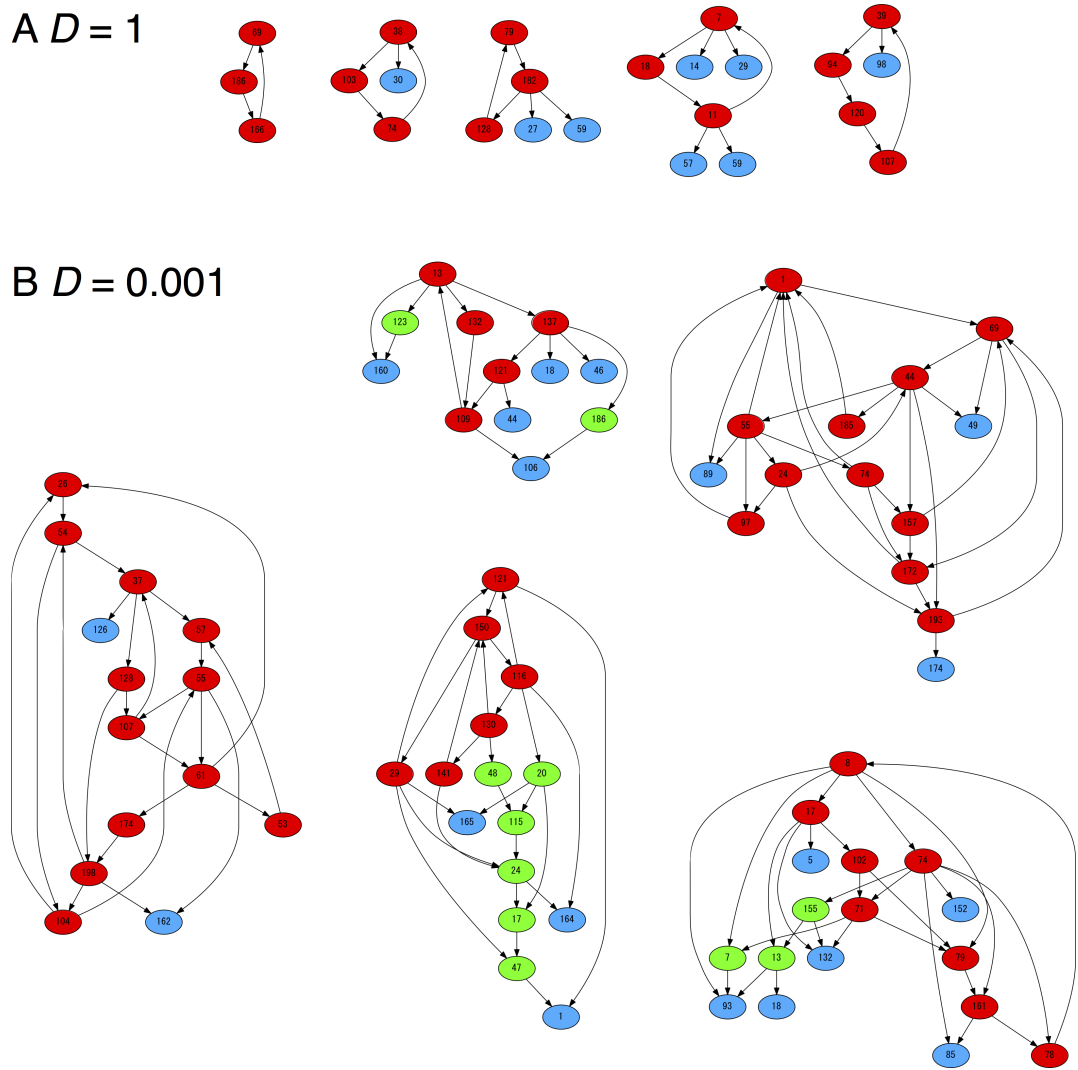}
\caption{The catalytic network formed by major molecule species. The major molecule species indicates that the copy number of the species is greater than ten. For each run of the simulations, almost all dividing cells have the same composition and adopt the identical network. 
For the parameters (\textbf{A}) D = 1 and (\textbf{B}) D = 0.001, such identical networks are shown obtained in five different underlying networks. The nodes indicate the molecule species. 
The arrows from $i$ to $j$ indicate the species $X_i$ catalyzes replication of $X_j$. 
By looking at the catalytic reaction network, the nodes are categorized into three types: host, sub-host and parasite molecule species. The host species (red) belong to at least one catalytic cycle so that the set of host species is auto-catalytic.  
Other than the host species, the sub-host species (green) indicates that they catalyze the replication of at least one other species in the major species (but does not belong to any autocatalytic hypercycle). The parasite species (blue) indicates that their replication is catalyzed by the other species, but they do not catalyze the other in turn. Parameters are $N_C = 100$, $N = 1000$, $K_M = 200$, $\mu = 0.01$.}
\label{fig2}
\end{figure}

When the cells uptake resources at a sufficiently rapid rate (e.g., for $D = 1$), three components typically dominate most of the composition for $N = 1000$ (each representing approximately 1/3 of the molecule population). The three components, say, $X_1$, $X_2$ and $X_3$, configure a catalytic cycle such that $X_1 \rightarrow X_2 \rightarrow X_3 \rightarrow X_1$, where $X_i \rightarrow X_j$ means replication of $X_j$ is catalyzed by $X_i$.
This catalytic cycle warrants that each of the species has a catalyst for its own replication. 
Since we excluded the direct mutual catalysis between $i$ and $j$, this three-component hypercycle\cite{eigenhypercycle} is a minimum auto-catalytic set (red nodes in Figure \ref{fig2}A). 
The hypercycle establishes a recursively growing state, where the composition is robust against stochasticity in the reactions and the division events. Most of the other molecule species are absent, while some species can appear by replication error from time to time. Some parasitic species could increase their number on occasion (blue nodes in Figure \ref{fig2}A). They are catalyzed by a member of the hypercycle but do not catalyze other members. However, cells dominated by the parasitic molecules cannot continue growth\footnote{In our model, any molecule species is not junk because it works as a catalyst with the counterpart molecule in Eq. (\ref{reaction}). However, the molecule species is a parasite if no counterpart molecule is present to be catalyzed. We will return to this point in section \ref{Result3}.}. Hence, those cells will be eliminated by selection at a cell level. All dividing cells adopt this three-component hypercycle, and there is no compositional diversity; cells use the minimum reaction pathway to grow.

By the cell-level selection shown in Figure \ref{fig1}B, cells with a faster growth rate outcompete those with a slower one. 
The Darwinian selection occurs because the resources are abundant. 
By regarding a set of $\{ X_i \}$ as a replicating entity $R_i$ in the argument of section \ref{Result1}, cells consisting of the fastest-replicating set of $\{ X_i \}$ will take over. 
The growth rate of cells is defined by the replicating rates of molecules. 
With the abundant resources, the replication rate of each molecule is determined by the product of concentrations of the reactants.
This product increases as the chemical abundance is concentrated on a fewer molecule species. 
Hence, cells without other components have a higher growth rate. Thus, the three-component auto-catalytic hypercycle dominates as a result of selection at a cell level. 

As $D$ decreases below a certain threshold $D_c \approx 0.01$, however, the number of molecule species increases, where multiple reaction pathways are utilized. As in the three-component hypercycle, the molecule species in this case also form a mutually catalytic hypercycle [red nodes in Figure \ref{fig2}B]. The other molecule species are connected to the species in the auto-catalytic set to replicate themselves [green and blue nodes in Figure \ref{fig2}B]. All dividing cells have approximately the same compositions. On occasion, cells dominated by parasites appear, but they cannot survive.

With limited resources, cells diversify their content to increase the growth rate. In this case, each replication rate is basically limited by the supply rate of its resource. Thus, cells grow fast if they utilize more variety of resources for their own growth. With the catalytic reaction network, cells with diverse molecule species can convert more variety of resources to replicate molecule species. Therefore, cells with diverse molecule species can outcompete simpler cells.

Hence, the diversification in cellular composition is a result of resource limitation. Here whether the resources are limited or not is determined by two relevant timescales, consumption and supply rates of the resource $S_i$ for replicating $X_i$. The consumption rate is inherently proportional to the product of concentrations of $X_i$ and its catalyst $X_j$. Thus, this consumption rate decreases as the number of intracellular molecule species increases. 
On the other hand, the maximum supply rate is given by a constant $DS_j^0$. The relative magnitude of the two timescales determines degree of resource limitation.

\begin{figure}
\centering
\includegraphics[width=8cm]{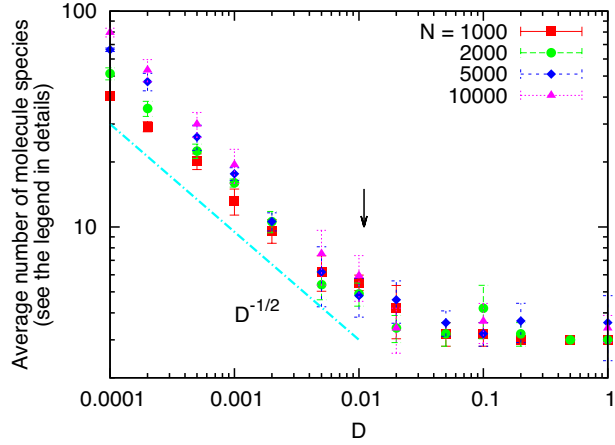}
\caption{The average number of molecule species (which have counterparts both as a catalyst and a template at division) as a function of 
$D$ for $N = 1000, 2000, 5000$, and $10000$. To eliminate the species that exist only as a result of mutation errors, we cut off the mutation errors after $2 \times 10^4$ division events and then count the number of species after the cells repeated $10^5$ divisions. The numbers have little dependence on N: they are approximately equal to 3, i.e., the minimum hypercycle for $D > D_c = 0.011$ (indicated by an arrow), 
where $D_c$ is estimated as the balance point between the maximum inflow and consumption rates of resources for the three-components hypercycle. 
Below $D \approx D_c$, the numbers increase as D decreases. 
The deviation in the range below $D =0.001$ for different $N$ occurs mainly due to the loss of some molecule species by division processes of the cells, in particular for $N = 1000$ and $2000$, because as the diversity increases, each molecule species decreases its number to several tens of copies at the division events. 
The slopes $D^{-1/2}$ is also shown for reference.}
\label{fig3}
\end{figure}

To balance the amounts of consumption and supply of resources, the consumption rate should decrease with the supply rate. 
When the consumption rate for the three-component hypercycle exceeds the supply rate, the molecular diversity starts to increase beyond three. 
This condition gives a transition point for $D$ to diversification [see Figure \ref{fig3}]. 
With the further decrease in supply below the critical point $D_c$, the optimal number of species for cell growth is expected to increase, as studied previously\cite{KamimuraKaneko2016}. 

Below this optimum number, the consumption rate is faster than the supply rate. Thus, resources are limited. Cells tend to diversify their contents to utilize more variety of resources for their own growth. 
Above the optimum number, the consumption rate is slower than the supply rate. Thus, the resources are abundant for the cell. Then, cells tend to simplify their content to increase the growth rate. 
Therefore, given the supply rate of resources, one expects the existence of an optimum number of molecule species to maximize the growth rate. 

From this optimization for growth, one can expect a quantitative relation between the number of molecule species and the supply rate. 
In fact, we found numerically that the number of molecule species increases with the parameter $D$ as $D^{-\alpha}$$(\alpha \approx 0.5)$ [see Figure \ref{fig3}]. 
This negative scaling relation is theoretically derived as follows. Let us denote the number of molecule species inside a cell by $K_M^*$$(0 < K_M^* \leq K_M)$.
In the catalytic reactions, the consumption of resource $S_i$ is written as $\approx S_i \rho_i \rho_j$. 
Because the concentration $\rho_i$ is approximately written as $\approx 1/K_M^*$, we obtain 
\begin{align}
    \frac{dS_i}{dt} \approx S_i/K_M^{*2} + D_r (S^0_i - S_i). 
    \label{r3}
\end{align}
Thus, the steady state condition of Eq. (\ref{r3}) is obtained as $\bar{S}_i = D_r S_i^0/(1/K_M^{*2}+D_r)$.
As the growth rate of the cells, $G$, is defined by the sum of replicating rates of molecules, it is approximately written as
\begin{align}
G = \sum_i \frac{d\rho_i}{dt} \approx \sum_i \frac{D_rS_i^0}{1 + D_r K_M^{*2}} \approx \frac{K_M^* D S^0}{1+ D K_M^{*2}},
\label{r4}
\end{align}
where $S^0$ denotes a typical value of $S_i^0$. The value of optimum number of species $K_M^{opt}$ to give the maximum of $G$ is obtained by $dG/dK_M^* = 0$. 
Thus, from Eq. (\ref{r4}), one gets $K_M^{opt} = D^{-1/2}$. Hence, the exponent $-1/2$ results from the second-order reactions of $X_i$ and $X_j$.
The exponent changes with the order of reactions, and may also depend on the structure of the reaction network (see \cite{KamimuraKaneko2016}).

We emphasize here that the negative scaling relation is obtained as a result of the multi-level selection between molecular replication and cellular growth. 
At a molecule level, the coexistence of various species is possible when the resources are limited.
However, the argument itself does not claim that the system prefers diversification, as molecule species with a higher replication rate has higher fitness at a molecular level. 
The diversification of molecule species occurs as a result of the selection pressure at the cell population level. 
The selection for growth speed causes cells to increase their components leading to the optimum level of diversity.

\section{Evolutionary constraints of the catalytic reaction network}
\label{Result3}

\subsection{The number of species is essential for high growth rate}

Cells with diverse species can increase their growth rates by utilizing a more variety of resources. This mechanism explains the advantage of increasing the number of molecule species. However, how the catalytic network expands its diversity over generations is not fully explored in our previous publication\cite{KamimuraKaneko2016}. 

Before investigating the process of diversification, we first note that the number of molecule species is essential to convert a more variety of resources for cell growth. The network structure itself is not relevant to the growth rate as far as every species has a catalyst for its replication. 

When the resources are abundant, cells prefer to simplify their components for their growth rate. Thus, the catalytic cycle is typically composed of the minimum number of species of three to four.  If there are multiple disjoint cycles, the cells select only a single cycle with the highest replication rate and exclude the others. Accordingly, all the nodes are connected as a single network.  

\begin{figure}
\centering
\includegraphics[width=\textwidth]{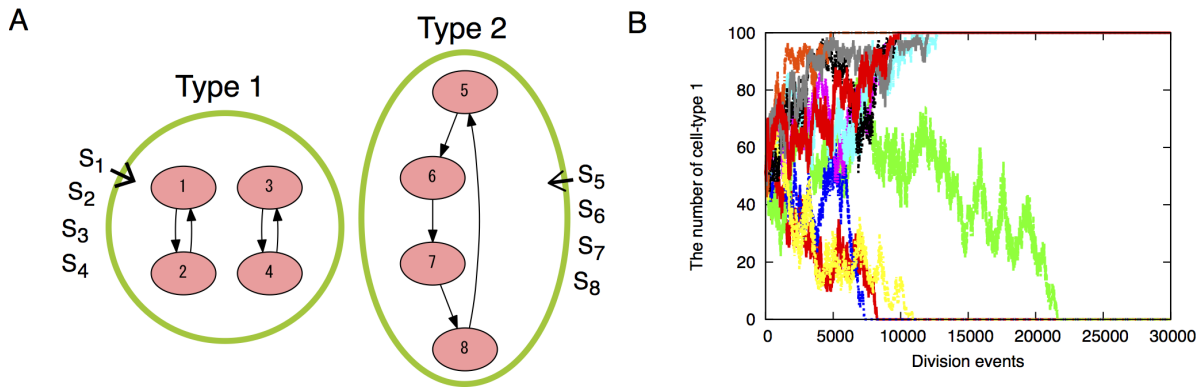}
\caption{ (\textbf{A}) Schematic representation of the simple model. We consider two types of reaction networks, of which each contains four molecule species. In type 1, four species $X_1$ to $X_4$ undergo the following reactions: 
$X_1 + X_2 + S_1 \rightarrow 2X_1+ X_2$, $X_2 + X_1 + S_2 \rightarrow 2X_2+ X_1$, $X_3 + X_4 + S_3 \rightarrow 2X_3+ X_4$, $X_4 + X_3 + S_4 \rightarrow 2X_4+ X_3$. In type 2, four species $X_5$ to $X_8$ undergo  $X_5 + X_8 + S_5 \rightarrow 2X_5+ X_8$, $X_6 + X_5 + S_6 \rightarrow 2X_6+ X_5$, $X_7 + X_6 + S_7 \rightarrow 2X_7+ X_6$, $X_8 + X_7 + S_8 \rightarrow 2X_8+ X_7$.
 (\textbf{B}) Competition of the two types of cells. Either of the two cells dominate the whole population. There is no selective advantage for either of the two, and one of the two types remains by chance for each run. Different colors indicate different simulation runs. In ten independent runs, type 1 dominates the population in 6 runs and type 2 dominates in 4 runs. Parameters are $N_C = 100$, $N = 1000$, $D = 0.001$, $c_i = 1$, $S_i^0 = 10$ for $i = 1,..,8$. }
\label{fig4}
\end{figure}

When the resources are limited, a single connected network itself is not essential for high growth rate. To demonstrate this, we consider a simple model in Figure \ref{fig4}A. We consider two types of cells. Both types are composed of four molecule species. In type 1, the molecule species form two sets of mutually catalytic cycles. 
In type 2, the molecule species form a single cycle. The other parameters are identical for the two types. Thus, type 1 has two independent cycles and type 2 has a single network. 
For this simple illustration, the direct mutual catalytic relation ($i \rightarrow j$, $j \rightarrow i$) is allowed here, instead of the three-component loop in the previous section. The same argument as presented here is applied for a comparison between two disconnected three-component hypercycles and one six-component loop.

The growth rates of the two types are equal when the resources are limited. Because the number of molecule species is essential for the growth rate, the structure of the catalytic network is not relevant. Thus, there is no selective advantage for a single joint network. In fact, survival of the cell types is by chance in direct competition between the two types [Figure \ref{fig4}B]. 
Even the stochastic reactions result in the dominance of the population by either type, no difference is observed in selection preference between the two types.  
The result of this simple model suggests that the joint network is not an absolute requirement for the higher growth rate.

\subsection{Cells diversify their molecule species by adding species one by one to the existing network}

Even though the single joint network is not an absolute requirement for the growth rate, 
it is generally observed as a result of diversification in which all the nodes (species) are connected [Figure \ref{fig2}B].  
We argue here that the joint network is obtained as an outcome of the evolutionary pathway to diversify their molecule species. 
In our simulation, a novel molecule species appears by errors in replication (``mutation"). The appearance of the new species by error is not sufficient for it to be fixed. The species has to increase its copy number. Otherwise, the species is diluted out by the growth of the cell. 

To successfully increase its copy number, the new species should have its catalyst in the cell. The fixation of the new species, then, is possible if the remaining species can catalyze this new species. Thus, the catalytic network diversifies its molecule species so that it connects the new species to the existing catalytic network. 

To attest this, we perform a simulation when the resources are limited ($D = 0.001$). 
In the initial condition of the simulation, each of the $N_C$ cells has only three molecule species $X_1$, $X_2$ and $X_3$.
The three species form the minimum hypercycle: $X_1 + X_3 \rightarrow 2X_1 + X_3$, $X_2 + X_1 \rightarrow 2X_2 + X_1$, and $X_3 + X_2 \rightarrow 2X_3 + X_2$.

\begin{figure}
\centering
\includegraphics[width=12cm]{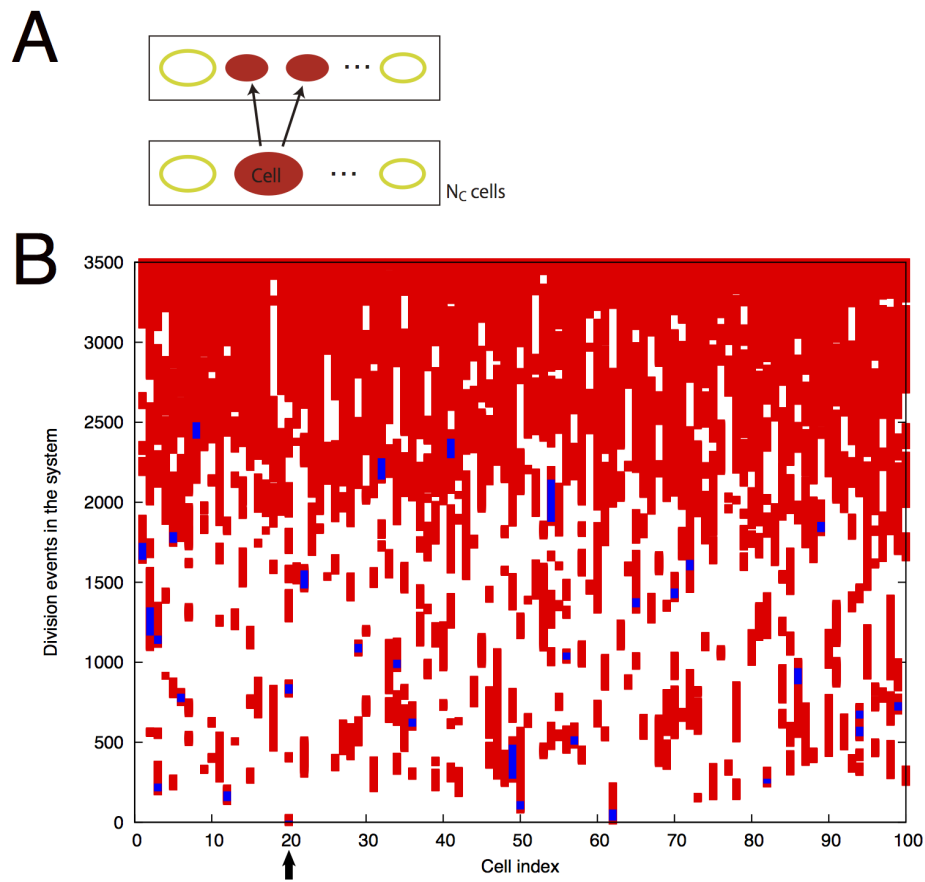}
\caption{ (\textbf{A}) The content of the cell is taken over by the two daughter cells. (\textbf{B}) By coloring the cells in red, one can see that all the $N_C$ cells are originated from one cell in the initial condition. In this sample, the initial cell at the position $20$ (indicated by the arrow) is the ancestor cell. We investigate how the number of molecule species increases by tracing the content of the progeny cells (up to the $2500$ division events; colored in blue) from the ancestor cell. 
Parameters are $N_C = 100$, $N = 1000$, and $D = 0.001$.
}
\label{fig5}
\end{figure}

From the minimum hypercycle, we trace the content of cells as shown in Figure \ref{fig5} to see how the cells diversify their molecule species. 
At cell division, the content of a cell is taken over by the two daughter cells [Figure \ref{fig5}A]. 
By coloring the daughter cells in red, we identify a single ancestor cell in the initial condition from which all the $N_C$ cells at the final stage are originated. 
In Figure \ref{fig5}B, all the cells are marked as red by the division events $3500$. 
Here we also trace the content of cells along a branch of such progeny cells from the ancestor cell [up to the $2500$ division events; colored blue in Figure \ref{fig5}B].

\begin{figure}
\centering
\includegraphics[width=12cm]{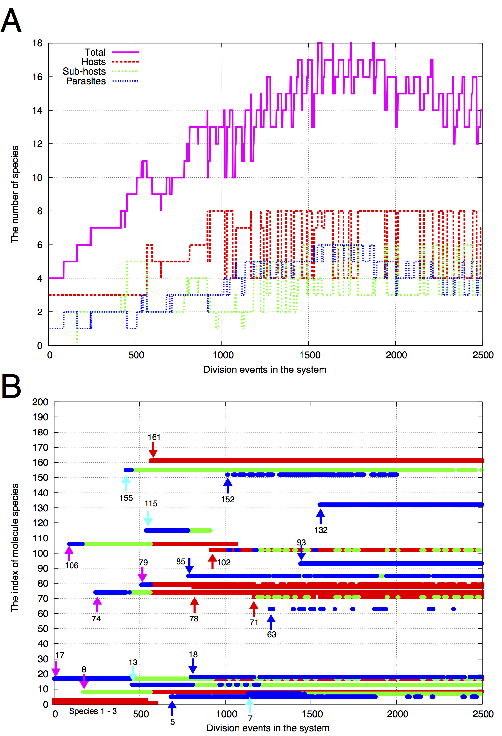}
\caption{ (\textbf{A}) The number of molecule species in the blue cells in Figure \ref{fig5}(B). Other than the initial three species ($1$, $2$, $3$; here we denote the species by its number $i$, instead of $X_i$), the number of the major molecule species (its copy number of the species is greater than ten) in each cell is plotted against the division events in the system. The total species (magenta) indicates the total of host (red), sub-host (green) and parasite (blue) species. See Figure \ref{fig2} for the definition of the species.   
(\textbf{B}) The indices of major molecule species in (A) are displayed for each cell at the corresponding time. The color of the points, red, green or blue, indicates a host, a sub-host or a parasite, respectively. Initially, the three species $1$, $2$ and $3$ are present in the cell. Other than the three species, twenty species appear and are fixed as the major species, whereas some of them are lost. The arrows indicate the time when such species appear. The red (blue) arrow indicates that the species appears and remains as a host (parasite) in the time period. On the other hand, the magenta arrow indicates that the species that initially appeared as a parasite (or a sub-host) changes its role to a host species (due to the appearance of other species). In addition, the light-blue arrow indicates that the species that appeared as a parasite changes its role to a sub-host species.   
}
\label{fig6}
\end{figure}

Figure \ref{fig6}A shows the number of major species in the cells along the branch. 
Other than the initial three molecule species ($X_1$, $X_2$, $X_3$), 
the major species is defined such that its copy number is greater than ten. 
The total species (magenta) indicates the number of such species. 
It increases from the beginning and eventually reaches a steady-state value ($\approx 15$ of this value of $D$). 
By looking at the catalytic network formed by the major species, we also show the number of host (red), sub-host (green) and parasite (blue) species. The host species indicate the member of an auto-catalytic hypercycle. 
Other than the host species, the sub-host species are defined such that they catalyze the replication of 
at least one other species in the major species, but do not belong to any auto-catalytic hypercycle.
The parasite species indicates that their replication is catalyzed by other species, but they do not catalyze any other in turn.

Furthermore, the major molecule species in a cell are displayed in Figure \ref{fig6}B. 
Initially, the three species $1$, $2$ and $3$ are present (hereafter, we denote the species by its number $i$, instead of $X_i$) and form a minimum hypercycle. Thus, the three species work as hosts and they are marked by red points. 
Shortly thereafter, the species $17$ appears as a parasite (with a blue point).
Then, the species $106$ appears also as a parasite.
The third species $8$ is fixed as a sub-host (with a green point) as it catalyzes the replication of $17$.
Simultaneously, the species $106$, originally a parasite species, changes its role to a sub-host because it catalyzes the replication of $8$ [the color of the points at species $106$ changes from blue to green by the appearance of the species $8$ in Figure \ref{fig6}B].

As the diversity increases by fixing parasite and sub-host species, a change of host species occurs. 
By the emergence of species $161$, several species change their role to host species (around 500 division events).
Even though the initial three species are almost simultaneously lost from the cells, the number of host species increases by successive transformation to host species from parasites. Then, most of the new species afterward are fixed and keep their role [shown with red and blue arrows], whereas some of them can be lost. 

At the initial stage, all the new species start as a parasite or a sub-host species (as shown with magenta and light-blue arrows). 
In fact, most of the new species initially emerged as parasites.
To start as a host species, the new species has to catalyze replication of the host species that exist. 
The diversity of the host species, however, is initially low so that the probability that the new species catalyze the host species is quite low. 
Hence, the new species has to start as a parasite species to the existing host species, or a parasite to a sub-host species, i.e., a parasite to the original parasite species.

One can roughly estimate the probability with which a new species can be initially introduced as a host species. 
The catalytic reaction path is assigned with probability $p$ (which was fixed at 0.1) for each pair of $X_i$ and $X_j$. Thus, the new species can catalyze one of the remaining host species with a probability $p/2$ on average (the factor 1/2 is added because only one of the two reactants works as a catalyst). 
By denoting the number of remaining host species by $K_H$, the probability that the new species can catalyze at least one host species is estimated as $p K_H /2$. 

After the fixation of species $161$ in Figure \ref{fig6}B, the number of host species ($K_H$) is approximately 5 or 6 between the division events $600$ and $900$. 
Thus, the probability $p K_H/2$ is estimated as 0.25-0.3. 
In fact, one species ($78$) is introduced as a host whereas three species ($5$, $85$, $18$) are introduced as parasites. 
Thereafter, the fixation of $102$ further increases the number of host species to eight, which increases the probability of appearance of host species later. 

\begin{figure}
\centering
\includegraphics[width=\textwidth]{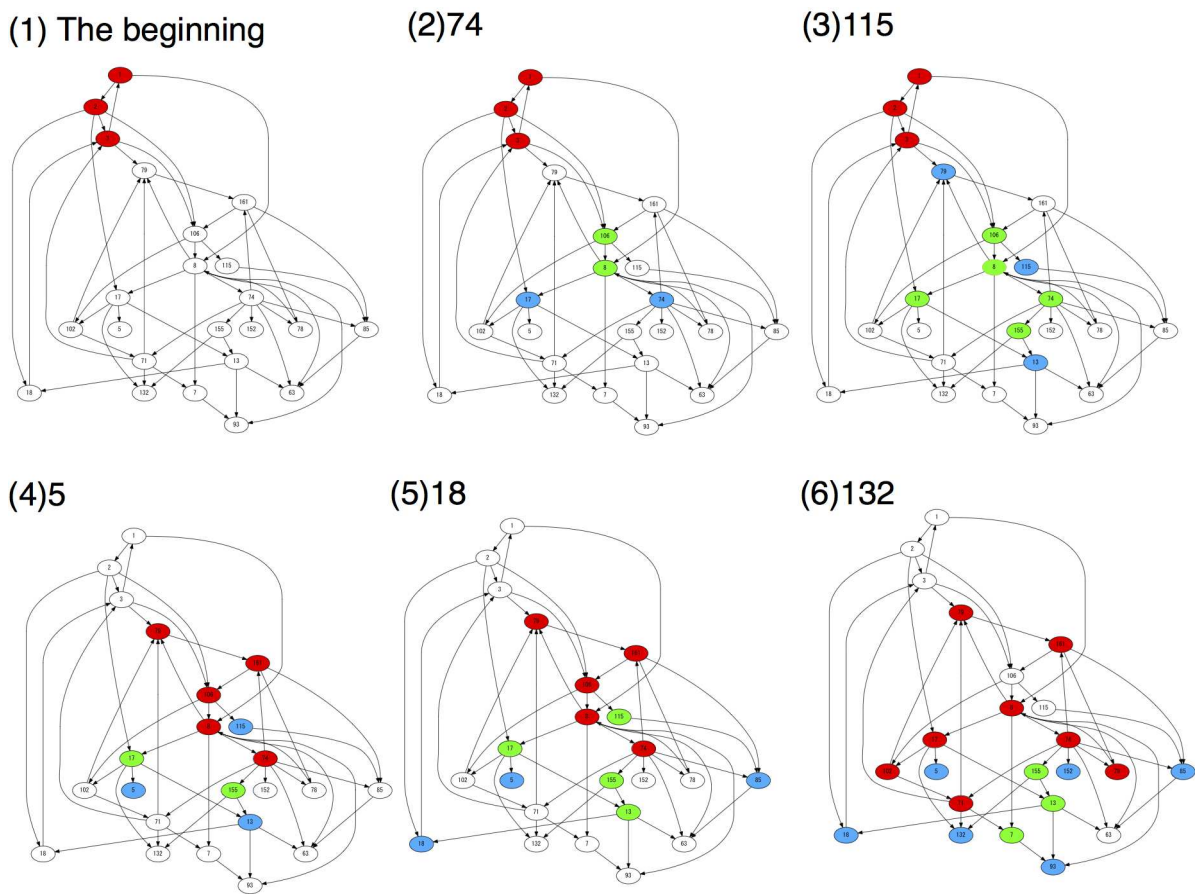}
\caption{The effective catalytic network is shown at the fixation of several molecule species in Figure \ref{fig6}B. The red, green and blue nodes correspond to the host, sub-host, and parasite species, respectively. The absent species at each time are represented by white nodes.
For (2) to (6), the number indicates the index of the newly fixed molecule species. 
}
\label{fig7}
\end{figure}

To further visualize the process of diversification, we show the effective catalytic network by coloring the existing molecule species in Figure \ref{fig7}. 
The underlying catalytic network is formed by the total of potential 23 molecule species, which appeared at some generations in Figure \ref{fig6}B. 
Here, the absent species at each generation are represented by white nodes. 

As we explain above, the new species initially appear and work as parasite or sub-host species [Figure \ref{fig7}(1) to (3)].
As the diversity increases, then, several species turn to be host species and a change of the ``core" network occurs [(4)]. 
With the successive increase of host species, the molecule species further diversify and a complex joint network evolves [(5) to (6)].

From this example, one can see that the cells fix their new species one by one to meet the requirements of growth and diversification simultaneously. 
This evolutionary constraint suggests a potential of ``parasitic" molecules. Typically, such species are considered as cheaters because they are not beneficial for maintaining the ``core" network. 
On the other hand, the species would be considered as a steppingstone toward diversification when the resources are limited. 

Here, the emergence of a new species is less plausible in a disjoint network. Such fixation requires construction of another catalytic cycle from scratch. To keep the number of the new molecule against the decreases by cell dilution, another mutation is necessary that catalyzes it, which hardly occurs.  Then, the mechanism of connecting the species is more plausible than constructing auto-catalytic cycle from scratch. 
In other words, the single connected network is more evolutionary achievable than the disjoint ones, even if the fitness (growth rate) is identical.

\section{Discussion}
\label{discussion}


Cells, in general, involve a huge variety of chemicals. For a given environment, in contrast, those with 
few components can grow faster. To resolve this apparent contradiction, we investigate how the cellular composition diversifies in a cell model consisting of the catalytic reaction network. 
As the resources are limited, the number of coexisting molecule species increases with which a variety of resources is converted to keep their growth.
Evolutionarily the diversity is increased by the appearance of parasitic species first and then parasites to the parasitic species.
Later they turn to be host species with further acquisition of novel molecule species. 

Our model assumes that each molecule species $X_i$ is replicated by consuming each resource species $S_i$.
This diversity of resource species in the environment is the underlying basis for the diversity of cellular components.
In this sense, the model is similar to the GARD model\cite{GARD}.
The GARD model is a kinetic model for homeostatic-growth and fission of an assembly of compositional lipids. 
It assumes biased accretion kinetics of molecular assemblies in diverse environmental molecules. 
In the growth of this assembly, the information of the composition (different types and quantities of molecules within an assembly) is transferred throughout the generations.
It will be then important to study the present diversity transition and scaling relation also in the GARD model.
The present result suggests the increase in the compositional information under the resource limitation. 
Recall that the information encoded in the composition is different from that encoded in RNA as the combinatorial diversity of sequences. 
Still, it is interesting to note that, under a limited flow of monomer resources, the sequence of catalytic polymers increases their complexity as has been shown recently\cite{matsubara2018}.  

In the present cells, all the diverse resources are not directly provided from the environment.
Instead, most substrates for each metabolic reaction are given by components which are products of intracellular reactions. Typical bacteria only need a source of basic elements (Carbon, Nitrogen, Phosphorus, Sulfur, ...) for their growth. Although the resource species are few, they are often decomposed by catabolic reactions, leading to diverse internal components. 
Then, through multi-body reactions, complex metabolic reactions follow with anabolic reactions. 
It will be interesting to extend the present study to include such multi-body reactions of polymers, 
and to understand the relevance of complex metabolic reaction networks to survival under resource-limited condition. 

In addition to the limitation of resources, competition between cells also gives a driving force for diversification of cell types, which was reported elsewhere\cite{KamimuraKaneko2015}. 
Also, in this case, cells diversify their molecule species in order to use a variety of limited resources for their own growth. 
In this case, however, cells with different components can use less-competitive resources so that they can increase their population. 
As a result, different types of cells appear in which different sets of molecules form different catalytic networks, and they coexist in the cell population. 

The diversification of molecule species in our study may remind of the niche differentiation in the field of ecology\cite{chesson2000,tilman1982resource}. When species differentiate to specialize for each niche, their competition is relaxed, so that their coexistence is easier. There exists, however, one important difference. 
A cell is a unit for selection, whereas an ecosystem itself is not a unit for selection as it does not reproduce. 
The ecosystem does not have an explicit selection pressure to grow faster.
In the present study, in contrast, the diversification of molecules is a result of the multi-level selection favoring a higher growth of a cell and higher replication of molecules. 
The multi-level evolutionary pressure leads to the formation of complex joint networks. This will be a unique feature of a cell system with multi-level evolution\cite{hogeweg1994multilevel,kaneko2006life}.

\section{Materials and Methods}
\label{methods}

We carried out simulations as follows. We introduce discrete simulation steps. For each simulation step, we repeat the following procedures. For each cell $q$ $(q = 1,...,N_C)$, we choose two molecules from the cell. If the pair of molecules, $X_i$ and $X_j$, are a replicator ($X_i$) and a catalyst ($X_j$), the replication of $X_i$ occurs with the given probability ($c_j)$ if $S_i^q \leq 1$.  $S_i^q$ is a continuous variable denoting the amount of the resource to replicate $X_i$ in the cell $q$. When the replication occurs, we add a new molecule of $X_i$ into the cell. At the same time, we subtract one resource molecule of $S_i$ to make $S_i^q \rightarrow S_i^q - 1$. With a probability $\mu$, we add a new molecule of $X_l$ $(l \neq i; S_l^q \leq 1)$, instead of $X_i$. If the total number of molecules in a cell exceeds the threshold $N$, the cell divides into two cells. We distribute the contents randomly into the two cells. At the same time, we remove one cell to fix $N_C$. We update each $S_i^q$ to $S_i^q + D (S^0_i - S^q_i)$ $(i = 1,...,K_M)$. 







\vspace{6pt} 



\authorcontributions{Conceptualization, A.K. and K.K.; Methodology, A.K. and K.K.; Software, A.K.; Formal Analysis, A.K.; Investigation, A.K.; Writing-Original Draft Preparation, A.K.; Writing-Review and Editing, A.K. and K.K.; Visualization, A.K.; Supervision, K.K.; Funding acquisition, K.K.}

\funding{This research was supported by a Grant-in-Aid for Scientific Research (S) (15H05746) from the Japan Society for the Promotion of Science (JSPS).}


\conflictsofinterest{The authors declare no conflict of interest. The funders had no role in the design of the study; in the collection, analyses, or interpretation of data; in the writing of the manuscript, or in the decision to publish the results.} 

\end{document}